# Rigorous Signal Reconstruction in Terahertz Emission Spectroscopy

Wentao Zhang and Dmitry Turchinovich*

*Fakultät für Physik, Universität Bielefeld, Universitätsstr. 25, 33615 Bielefeld, Germany*

*\*dmtu@physik.uni-bielefeld.de*

*Abstract*

Terahertz (THz) emission spectroscopy is a powerful method that allows one to measure the ultrafast dynamics of polarization, current, or magnetization in a material based on THz emission from the material. However, the practical implementation of this method can be challenging, and can result in significant errors in the reconstruction of the quantity of interest. Here, we experimentally and theoretically demonstrate a rigorous method of signal reconstruction in THz emission spectroscopy, and describe the main experimental and theoretical sources of reconstruction error. We identify the linear line-of-sight geometry of the THz emission spectrometer as the optimal configuration for accurate, fully calibrated THz signal reconstruction. As an example, we apply our reconstruction method to ultrafast THz magnetometry experiment, where we recover the ultrafast magnetization dynamics in a photoexcited iron film, including both its temporal shape and absolute magnitude.

*Introduction*

Terahertz (THz) emission spectroscopy is a powerful method of measurement of ultrafast dynamics of polarization, conduction current, or magnetization in the material subject to ultrafast excitation. This method is based on the fundamentals of electrodynamics: transient change in the polarization, current or magnetization becomes the source of free-space electromagnetic radiation [1], which can be measured and ultimately traced back to the dynamics of its source. If the timescale of the transient change is picosecond or sub-picosecond, then the emitted free-space radiation belongs to the THz spectral range. So far, THz emission spectroscopy has been widely applied to the studies of bulk and nanostructured classical and organic semiconductors [2–7], superconductors [8], graphene [9,10], and magnetic systems [11–17].

The essential challenge in successful implementation of THz emission spectroscopy is the rigorous reconstruction of the ultrafast dynamics of the quantity of interest: polarization, current or magnetization, serving as the source of THz emission, from the observable, the actually measured THz signal. In contrast to standard THz time-domain spectroscopy [18–20], which in its standard implementation is a self-referenced method and does not require the precise knowledge of the experimental setup architecture, in THz emission spectroscopy the details of the setup geometry and particulars of the THz signal detection scheme play a crucial role in accurate signal reconstruction. The experimental and theoretical studies of generation, propagation and detection of ultrafast THz signals in experimental setups have been widely reported in literature, see e.g. [21–27]. However, these studies mainly focused on the general transformation of the THz waveform during the generation, propagation, and detection, whereas the comparison between the theoretical and experimental data was presented only qualitatively. While these results significantly help us to understand the factors connecting the source of the THz emission to the actually measured THz signal, the state-of-the-art knowledge is not yet sufficient for the mathematically rigorous and fully calibrated implementation of THz emission spectroscopy.

In this paper, we describe the processes in the generation, propagation and detection of the THz waveforms in the ultrafast THz emission experiment, and provide the calibrated complex-valued transfer function connecting the time-varying source, i.e. polarization, current or magnetization in the THz emitter, to the observable, the THz electro-optic (EO) signal detected in the far field. We calculate the THz field distribution at an arbitrary position of the beam path, as well as THz signal distortion during the detection via free-space electro-optic sampling (FEOS). In our work, we first describe the THz emission spectrometer based on the ZnTe crystals for both THz emission and detection (see e.g. [23] and references therein), and rigorously establish the complex-valued transfer function, permitting the accurate deconvolution of the measured THz EO signal back to its source – transient polarization in the emitter ZnTe crystal. We then use this transfer function in a calibrated reconstruction of ultrafast magnetization dynamics in an ultrafast THz magnetometry measurement, where the THz radiation is driven by the magnetic dipole emission from a laser-excited iron film, serving as a THz emitter. We also analyze the dominant sources of experimental error in successful implementation of THz emission spectroscopy experiment, and identify the linear line-of-sight THz emission spectrometer as an optimal configuration for such type of measurements.

*Experiment and discussion*

Here, we describe our experiment, in which the full transfer function of the linear line-of-sight THz emission spectrometer, schematically shown in Fig. 1, is characterized. A transform-limited 100 fs laser pulse at 800 nm central wavelength from an amplified Ti:sapphire laser is split into pump and probe beams. The collimated pump beam is normally incident on a <110> ZnTe crystal of 0.5 mm thickness, with a spot size of 19.6 mm² ($1/e^2$) and a pulse fluence of 51 µJ/cm², thus producing the THz radiation via optical rectification in ZnTe. The generated THz beam then passes through an optical blocker, which filters out the optical laser beam, and THz-transparent optical pellicle beamsplitter, and is finally detected via FEOS in another <110> ZnTe of 1 mm thickness. The distance between the two ZnTe crystals is 10 cm. This setup does not include any additional optics between the emitter and the detector, eliminating any possible influence of additional misalignment, and thus allowing us to accurately compare our experimental results with those of our modeling.

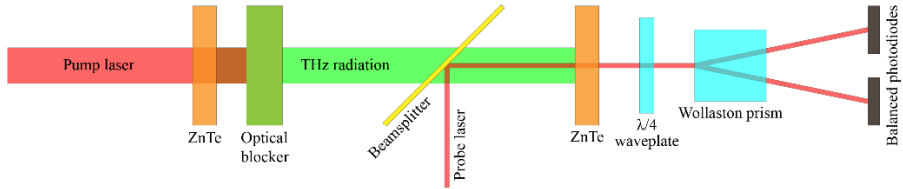

Fig. 1. Schematic of the THz experimental setup.

The THz generation in ZnTe is based on the optical rectification, a second order nonlinear effect. The femtosecond laser pulse induces a polarization in ZnTe which temporally follows its intensity envelope. This polarization then serves as a source of electric dipole radiation into the free space, thus producing a THz pulse. The laser-induced polarization can be written in the frequency domain as [23,28]

$$P_i(\omega) = \int_{-\infty}^{\infty} d\omega' \sum_{j,k} \varepsilon_0 \chi_{ijk}^{(2)}(\omega; \omega', -\omega' + \omega) E_j(\omega') E_k(\omega' - \omega), \tag{1}$$

where $\chi^{(2)}$ is the second order nonlinear susceptibility, and $E$ is the electric field strength of the optical laser pulse. Electric field in the frequency domain, under an assumption of Gaussian pulse shape, is $E(\omega' - \omega_0) = \text{FT}\{[0.94 F_p Z_0/(2n\tau)\exp(-2.77 t^2/\tau^2)]^{1/2}\}$, where $\omega_0$ is the central angular frequency, $F_p$ is the fluence, $Z_0 = 377$ Ohm is the vacuum impedance, $n$ is the refractive index, $t$ is time, $\tau$ is the pulse duration, and FT stands for Fourier transform. The time-domain polarization signal can be readily obtained by inverse Fourier transformation of Eq. (1). A current $J$ is connected to the polarization $P$ via $J(t) = \partial P/\partial t$.

From the far-field point source approximation, the electric field of the *electric dipole* THz emission can be obtained by [1]

$$\mathbf{E}_{\text{THz,p}} = \frac{1}{4\pi\varepsilon_0} k^2 (\mathbf{n} \times \mathbf{p}) \times \mathbf{n} \frac{e^{ikr}}{r}, \tag{2}$$

where $r$ is the distance between an arbitrary point in space and the emitting electric dipole, $k$ is the wave vector of the THz wave, $\mathbf{p}$ is the electric dipole moment induced by the optical laser (see Eq. (1)), and $\mathbf{n}$ is a unit vector pointing from the emitting electric dipole towards the point in space. We define a radiation function as $f_{\text{rad,p}}(\omega) = E_{\text{THz,p}}/P$, connecting the polarization and the emitted electric field. For completeness, we also present here an analogous expression for the *magnetic dipole* emission driven by the magnetic dipole moment $\mathbf{m}$, also in a point source approximation [1]

$$\mathbf{E}_{\text{THz,m}} = -\frac{1}{4\pi}\sqrt{\frac{\mu_0}{\varepsilon_0}} k^2 (\mathbf{n} \times \mathbf{m}) \frac{e^{ikr}}{r}. \tag{3}$$

We note, that the polarization in a laser-excited ZnTe crystal cannot be universally regarded as a point source, since the diameter of the laser spot on the crystal is usually of the order of several millimeters, thus exceeding the generated wavelength, unless the detection position is sufficiently far enough from the source. In order to apply Eqs. (2) and (3), we discretize the extended polarization or magnetization source into small pieces, which is then described as an array of point sources. Finally, the electric field at arbitrary point in space is obtained numerically from the superposition of the contributions of individual point sources.

The phase velocity of the generated THz wave $v_{\text{THz}}$ differs from the group velocity of the optical pulse $v_g$ in the ZnTe crystal, thus limiting the generating efficiency and the bandwidth of the THz emission. The normalized frequency-dependent filter function for the phase-matching between the optical and THz pulses within the crystal is given by [26]

$$f_{\text{PM}}(\omega) = \frac{e^{i(n_{\text{THz}} - n_g)\omega L/c} - 1}{i(n_{\text{THz}} - n_g)\omega L/c}, \qquad (4)$$

where, $n_{\text{THz}}$ is the THz refractive index of the generating crystal at angular frequency $\omega$, $n_g$ is the group refractive index of the crystal for the optical laser pulse, and $L$ is the thickness of the crystal. Note here that, in the phase-matching term, the THz refractive index can have a complex value, also accounting for the THz absorption within the crystal. Therefore, Eq. (4) describes not only the THz-to-optical phase-matching, but also the THz dispersion and absorption within the crystal.

The last effect taken into account in the THz generation is the Fabry–Perot multi-reflection inside the ZnTe crystal, which becomes more prominent for the thinner crystals. The Fabry–Perot term is [21]

$$f_{\text{FP}}(\omega) = \frac{t_{10}}{1 - r_{10}^2 e^{2in\omega L/c}}, \qquad (5)$$

where $r_{10}$ and $t_{10}$ are the THz field reflection and transmission Fresnel coefficients from the crystal to air, respectively.

The FEOS detection in a ZnTe crystal [23–27] is based on the THz-induced phase retardation of different polarization components of the initially linearly polarized optical probe pulse within the crystal. This phase retardation is produced via the Pockels effect in ZnTe, induced by the co-propagating THz field, and in a small-signal approximation is directly proportional to the THz electric field. This approximation is valid for the THz fields not exceeding several 100s V/cm strength, which is typically the case in THz emission spectroscopy experiments. The THz-induced phase retardation of the optical probe beam is analyzed using a quarter-wave plate, a polarization beamsplitter, and a pair of balanced photodiodes located after the ZnTe detection crystal (see Fig. 1). The measured quantity in FEOS is the difference of the intensities of the illumination of the two photodiodes $\Delta I$, which, within a small signal approximation, is directly proportional to the instantaneous THz electric field $E_{\text{THz}}$ including its sign. Under the condition of normal incidence of the THz and the optical probe beams on a <110> ZnTe crystal, we have, following [25]

$$f_{\text{EO}}(\omega) = \frac{\Delta I/I}{E_{\text{THz}}} = t_{01} \frac{\omega_0 n_0^3 r_{41} L}{2c} (\cos\alpha \sin 2\varphi + 2\sin\alpha \cos 2\varphi), \qquad (6)$$

where $L$ is the thickness of the detecting crystal, $n_0$ is the refractive index of the optical probe pulse at central angular frequency $\omega_0$, $r_{41}$ is the nonzero coefficient of the EO tensor of the crystal, $I$ is the total intensity of the probe beam, $t_{01}$ is the THz field transmission Fresnel coefficient from air to the crystal, and $\alpha$ and $\varphi$ are the angles of the polarizations of the THz and the probe beams with respect to the <001> axis of the ZnTe crystal, respectively. In an experiment, since the duration of the probe laser pulse $\tau$ has a finite value, the THz field features of the duration shorter than $\tau$ are smeared out. In the frequency domain, this effect acts as a low-pass filter, narrowing the bandwidth of detected THz EO signal [23]:

$$f_\tau(\omega) = \exp\left(-\frac{\omega^2 \tau^2}{16 \ln 2}\right). \qquad (7)$$

It should be noted that Eqs. (1)-(7) are applicable not only to ZnTe, but to any other nonlinear/EO crystal, suitable for the THz generation via optical rectification and THz detection via FEOS in collinear geometry.

In Figs. 2(a)-2(c) we demonstrate, in the frequency domain, the influences of the factors mentioned above on the THz signal detected in a modelled spectrometer depicted in Fig. 1. The corresponding time-domain representations can be generated using the inverse Fourier transformation of the frequency-domain functions. The polarization $P(\omega)$, generated by the 100 fs, 800 nm laser pulse with the fluence of 51 µJ/cm² in the 0.5 mm ZnTe crystal (Eq. (1)) is shown in Fig. 2(a). The electric dipole radiation term $f_{\text{rad,p}}(\omega)$ is shown in two variants – in point source approximation (Eq. (2)) as a solid green line in Fig. 2(b), and taking into account the finite laser spot size and emitter-detector separation in our experiment as a solid black line in Fig.

2(b). Note the differences between these two curves, which demonstrate the importance of taking into account the realistic pump spot size in THz emission experiments. The following parameters were calculated using the known THz-range dielectric function of ZnTe, and optical group index $n_g = 3.239$ at 800 nm central wavelength [29]: the phase-matching term $f_{PM}(\omega)$ (Eq. (4)) in generation (black lines in Fig. 2(c)) and detection (red lines in Fig. 2(c)). As a solid green line in Fig. 2(c), we show the reduction of time resolution in THz FEOS due to the finite probe laser pulse duration $f_\tau(\omega)$ (Eq. (7)). We note that the EO coefficient $r_{41}$ for ZnTe is almost a constant within the spectral range of our experiment 0.0-3.0 THz [30].

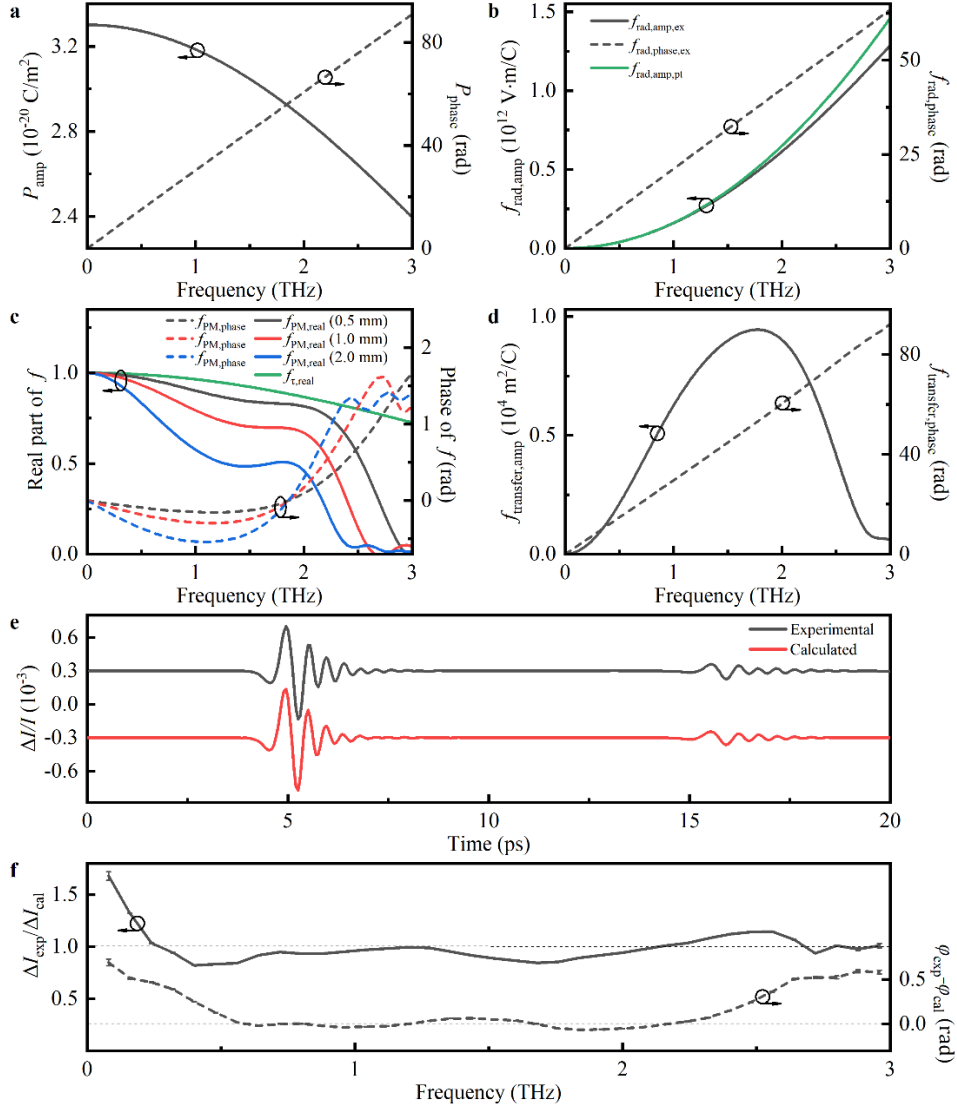

Fig. 2. Comparison of modelled and measured performance of a line-of-sight THz emission spectrometer: electric dipole radiation is generated in 0.5 mm – thick ZnTe crystal by 800 nm laser pulses of 100 fs duration with the fluence of 51 µJ/cm², and detected in a 1 mm – thick ZnTe crystal. (a) Polarization induced by the laser pulses. (b) Electric dipole radiation from extended and point sources. (c) Functions of phase-matching and nonzero pulse duration of the probe pulses. (d) Transfer function of the setup. (e) Calculated and experimental THz EO signals. (f) The discrepancy function obtained from $\Delta I_{exp}/\Delta I_{cal}$. The circled arrows in figures refer to the relevant axes.

Essentially, the bandwidth of the detected THz radiation is mainly limited by the phase-matching effect (Eq. (4)) at higher frequencies, and by the dipole radiation effect (see Eq. (2) for a point-source approximation) at lower frequencies. Additionally, we also calculated the phase-matching of 2 mm thickness (blue lines in Fig. 2(c)). The phase-matching is controlled effectively by the thickness of the crystal, the thinner the crystal, the larger the acceptance bandwidth. Further, we emphasize it again that the difference between the dipole radiation terms in the point-source approximation and for the realistic emitting spot size is found to be a non-negligible parameter in our experiment with the emitter-detector separation distance of 10 cm.

The detected time-domain EO signal $S(t)$ is obtained from the inverse Fourier transformation (iFT) of the product of the above terms, which is written as

$$S(t) = \text{iFT}\{P(\omega)f_{\text{rad,p}}(\omega)f_{\text{PM,e}}(\omega)f_{\text{FP}}(\omega)f_{\text{OB}}(\omega)f_{\text{BS}}(\omega)f_{\text{EO}}(\omega)f_{\text{PM,d}}(\omega)f_{\tau}(\omega)\}, \quad (8)$$

where $f_{\text{OB}}(\omega)$ and $f_{\text{BS}}(\omega)$ are the THz transmission functions of the optical blocker and the pellicle beamsplitter, respectively, which are easily obtained experimentally.

In Fig. 2(e) we show the calculated THz EO signal in time domain (Eq. (8)) together with the experimentally measured one. The main THz pulse and a small time-delayed reflection signal are observed. The time delay between the two signals is around 10.5 ps. The agreement between the fully modelled and the measured THz EO signals is very good. To quantify the agreement between the modelled and the measured signals, we determine the discrepancy function, presented in Fig. 2(f) in the frequency domain as the ratio of the Fourier transforms of the experimental and modelled THz pulses from Fig. 2(e). This discrepancy is rather small, and may arise from the non-ideality of the experimental conditions, such as the slight off-axis propagation of the optical and THz beams, and from the fact that the lens-focused EO sampling optical beam in our experiment was assumed to be an infinitely thin beam in our calculations. Otherwise, our calculated result reproduces the experimental data very well. It should be noted that the experimental and calculated THz EO signals presented in Fig. 2(e) are shown in absolute values (see Eq. (6)), i.e. without normalization or scaling of any kind.

The good agreement between the experimental and calculated THz EO signals produced via electric dipole THz emission from laser-excited ZnTe crystal indeed proves that our modeling can provide accurate quantitative description of the THz emission spectrometer.

The functions $f_{\text{OB}}(\omega)f_{\text{BS}}(\omega)$ and $f_{\text{EO}}(\omega)f_{\text{PM,d}}(\omega)f_{\tau}(\omega)$ corresponding to the propagation and detection of the THz signal in the spectrometer, are independent of the THz radiation source. Therefore, the function

$$f_{\text{transfer,p}}(\omega) = f_{\text{rad,p}}(\omega)f_{\text{OB}}(\omega)f_{\text{BS}}(\omega)f_{\text{EO}}(\omega)f_{\text{PM,d}}(\omega)f_{\tau}(\omega), \quad (9)$$

represents a fundamental transfer function of the THz emission spectrometer, connecting the THz emission source, electric dipole, with the experimental observable - its detected EO signal. This transfer function representing the experiment from Fig. 1 is shown in Fig. 2(d). We note here that the transfer function of the THz emission spectrometer can be established experimentally, as long as a well-calibrated reference emitter is employed, such as a ZnTe crystal as described here. The very small non-ideality factor, shown in Fig. 2(f), representing the deviation between the measured and fully simulated THz EO signals (see Fig. 2(e)), shows that the above theoretical description of the operation of a THz emission spectrometer is sufficient for rigorous reconstruction of signals with THz emission spectroscopy.

Depending on whether the electric or magnetic dipole radiation is being measured, the straightforward connection between the observable, the measured THz EO signal, and the quantity of interest, polarization $P$ or magnetization $M$ dynamics in the excited sample, can be expressed in the frequency domain as

$$P(\omega) = \frac{S(\omega)}{f_{\text{transfer,p}}(\omega)f_{\text{out}}(\omega)}, \quad (10)$$

$$M(\omega) = \frac{S(\omega)}{f_{\text{transfer,m}}(\omega)f_{\text{out}}(\omega)}, \quad (11)$$

taking into account different radiation mechanisms Eqs. (2) and (3), where $f_{\text{out}}(\omega)$ is the out-coupling function, depending on the THz source itself. For ZnTe crystal, $f_{\text{out}}(\omega) = f_{\text{PM,e}}(\omega)f_{\text{FP}}(\omega)$. We note that, since the generation terms $P(\omega)f_{\text{PM,e}}(\omega)f_{\text{FP}}(\omega)$ for ZnTe are well known, the ZnTe crystal serves as a good THz source for the emission spectrometer calibration.

Now, we apply the obtained spectrometer transfer function corresponding to magnetic dipole radiation and demonstrate the reconstruction of the magnetic dipole THz emission signal from the laser-excited iron film. In Fig. 3 we present the THz EO signal originating from magnetic dipole emission from 10 nm Fe film, deposited on MgO substrate of 0.5 mm thickness, capped with MgO layer of 12 nm thickness, and excited by the 800 nm laser pulses of 100 fs pulse duration, 0.51 mJ/cm² energy density, and 19.6 mm² ($1/e^2$) spot size; and the corresponding reconstruction of magnetization dynamics within the laser-excited Fe film. The THz

EO signal in this measurement was detected in a 1 mm – thick ZnTe crystal. The details of this experiment, as well as the physical mechanism behind the complex demagnetization process in laser-driven iron, are described in detail in [15]. Here, we present this result as a demonstration of the potential of the quantitative THz emission spectroscopy described in this paper, for accurate ultrafast magnetometry measurements.

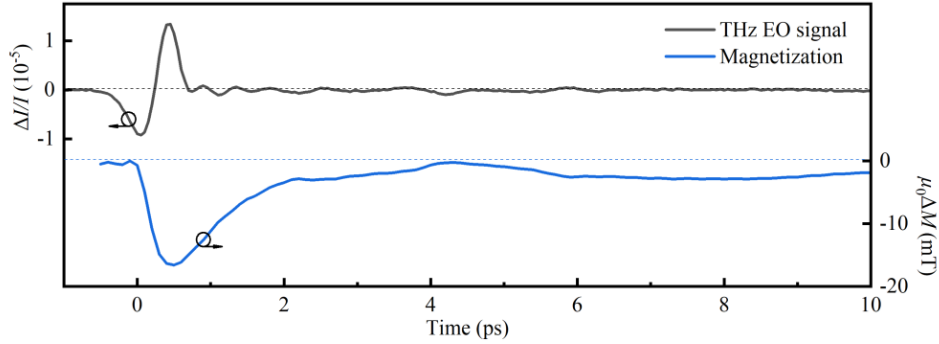

Fig. 3. Example of ultrafast THz magnetometry measurement using a line-of-sight THz emission spectrometer. Top curve: the THz EO signal originating from magnetic dipole emission from 10 nm Fe film, deposited on MgO substrate of 0.5 mm thickness, capped with MgO layer of 12 nm thickness, and excited by the 800 nm laser pulses of 100 fs pulse duration, 0.51 mJ/cm$^2$ energy density, and 19.6 mm$^2$ ($1/e^2$) spot size. The THz EO signal is detected in a 1 mm – thick ZnTe crystal. Bottom curve: the reconstructed magnetization dynamics induced by the laser excitation in the Fe film. The circled arrows in figure refer to the relevant axes.

*Simulation of THz propagation in a complex spectrometer setup, and effects of setup misalignment*

Here, we simulate the THz beam propagation in a relatively complex THz setup containing more optical elements, and analyze the effect of misalignment on the accuracy of signal reconstruction in a corresponding THz emission experiment. Commonly, in THz spectroscopy setups, off-axis parabolic mirrors (OAPMs) are used to focus and collimate the THz waves, in order to achieve a smaller THz beam spot at the sample position, and stronger THz field at the detection position. In our modeling, we use the Kirchhoff's diffraction formula to calculate the THz field distribution after the reflection of the OAPMs. At an arbitrary position $\mathbf{x}$, the field $U(\mathbf{x})$ from an extended source $U(\mathbf{x}')$ at surface $A$ is [31]

$$U(\mathbf{x}) = \frac{1}{4\pi} \int_A \left[ U(\mathbf{x}') \frac{\partial}{\partial n}\left(\frac{e^{ikr}}{r}\right) - \frac{e^{ikr}}{r} \frac{\partial U(\mathbf{x}')}{\partial n} \right] dA, \qquad (12)$$

where, $n$ is the normal direction at position $\mathbf{x}'$ on $A$, $r = |\mathbf{x} - \mathbf{x}'|$ is the distance between $\mathbf{x}'$ and $\mathbf{x}$. Here, $U(\mathbf{x})$ can be electric or magnetic field. Using Eq. (12), we calculate the generated field on a surface away from the source, which is then used as the secondary source in the calculation of the field on yet another surface. This sequence of calculation is performed iteratively until the target surface.

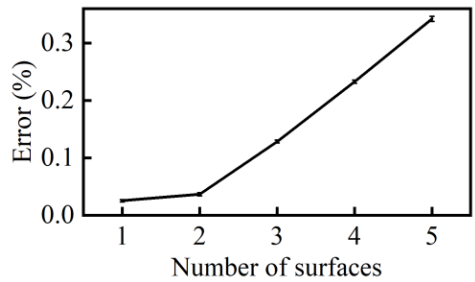

Fig. 4. Error of the calculated THz electric field as a function of the number of the surfaces between the source and the detecting surface, induced by the iterative calculation.

Firstly, we check the reliability of our simulation approach by solving Eq. (12) numerically, and estimate the error in the calculation, as the complexity of the beam path increases. Our error estimation is as follows. After forming the THz beam from an array of elementary electric dipoles in the emitter, we propagate this THz beam in space and time. Then, at a certain coordinate along the THz beam propagation we create an artificial trial surface, and determine the electric field distribution from the incident THz beam at this surface. This is equivalent to creating a new array of elementary THz-emitting sources of electric field, originating at this surface. We thus treat this trial surface as a new THz emitter, and form the new THz beam originating from it, which is then propagated further. Obviously, in an ideal case of a perfectly accurate calculation of THz propagation, the artificial creation of such trial surfaces within the THz beam path, which are then treated as secondary THz emitters, should not lead to any change in the propagation of the original THz beam. However, a numerical calculation can generate an error in propagation, which should grow with the number of such trial surfaces introduced. We therefore have verified the numerical accuracy of our THz propagation calculation by inserting 5 such trial intermediate surfaces in the beam path of a modelled linear line-of-sight spectrometer shown in Fig. 1. The error in the calculation of the THz electric field as a function of the number of inserted surfaces is shown in Fig. 4, with the maximum error reaching the value of ~0.35% after 5 trial surfaces. This error is negligibly small, thus confirming the accuracy of our numerical approach to treat the THz propagation.

Now, using Eq. (12) and the approach described above, we perform a calculation based on the spectrometer geometry shown in Fig. 5(a), which is a standard configuration for a THz time-domain spectrometer based on the ZnTe crystals, using intermediate focusing optics. In this calculation, we assumed that OAPM 1 has a diameter and reflected focal length (RFL) of 1 inch, and OAPMs 2-5 have 2-inch diameter and 4-inch RFL, respectively. A plane source of electric dipole emission with a Gaussian distribution of 8-mm diameter ($1/e$) generates the THz radiation. The diameter of the THz beam is magnified by OAPMs 1 and 2, and the beam is focused at the sample position and at the detection position by OAPMs 3 and 5, respectively. The THz field distributions at these two positions are calculated and their peak values are shown in Fig. 5(b). Basically, the THz electric field increases with frequency (blue and red solid lines), which follows from Eq. (2). For individual frequencies, the electric field and the spot diameter (blue and red dashed lines) remain nearly unchanged at both positions, which means that OAPMs 4 and 5 can image the THz field distribution of sample position accurately onto the detection position. However, due to the divergence effect, the beam diameter at the detection position is slightly larger, resulting in a slightly lower electric field.

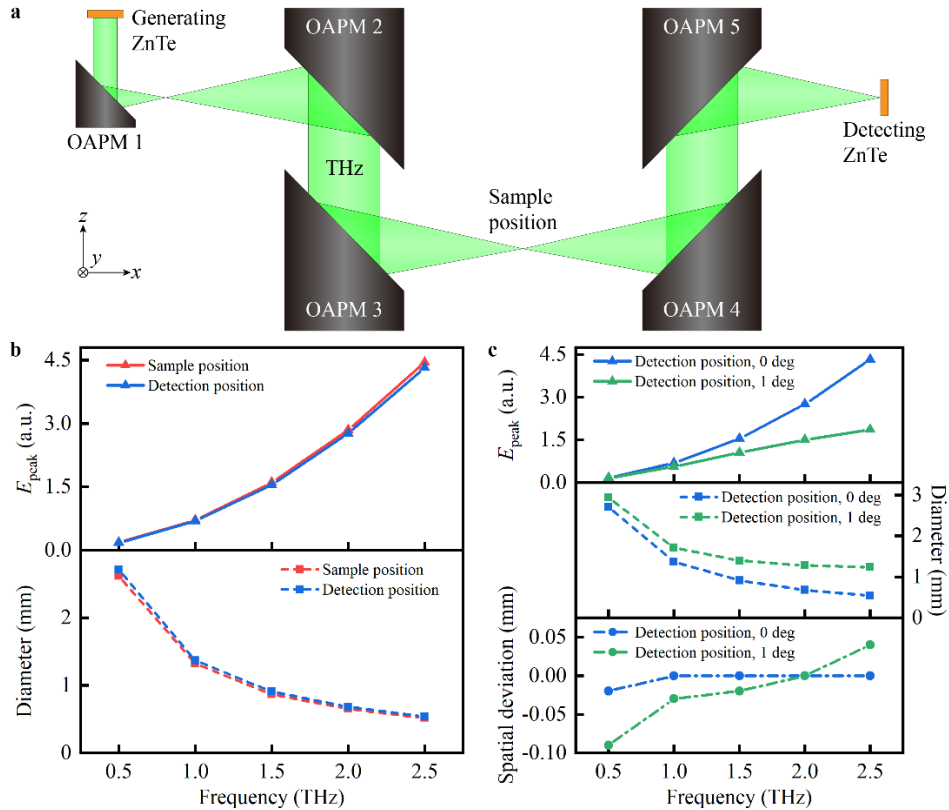

Fig. 5. (a) Schematic of the THz beam path in the calculation of the THz propagation. (b) Peak values of the THz electric field and the diameters of the beam spots at sample and detection positions. (c) Peak values of THz electric field, the diameters of the beam spots, and the spatial deviations of the peaks at detection position without misalignment (blue), and after OAPMs 4 and 5 are rotated clockwise by 1 degree (green).

Next, we simulate the setup misalignment: we rotated OAPMs 4 and 5 clockwise in *x-z* plane by 1 degree, to observe how this rather minor misalignment affects the THz field distribution at the detection position. In Fig.

5(c) we show the peak values of the THz electric field of different frequencies, the beam spot diameters, and the deviations of the peaks from the geometric focus of OAPM 5 without misalignment (blue), and with misalignment of OAPMs 4 and 5 by 1 degree (green). This rather minor misalignment of 1 degree leads to a significant field distortion at the detection position: the detected electric field becomes much smaller, especially for higher frequency components, the spots become larger, and also the peaks deviate slightly from their original positions. We note here, that the deviations of the peaks, caused by the setup misalignment, are rather minimal compared to the size of the beam spots. On the other hand, the misalignment-induced decrease of the electric field strength at the detection position is rather severe, with the field reduction factor at higher frequencies reaching the value of ~2.5. As a result, even a minor misalignment of 1 degree of the THz guiding optics in the spectrometer shown in Fig. 5(a) certainly leads to the complex and difficult to account for distortion of the detected THz EO signal, which will be impossible to correct in a precise signal reconstruction. Therefore, the direct line-of-sight configuration of the THz emission spectrometer, shown in Fig. 1, in spite of the lack of THz focusing and hence the lower overall detected EO signal strength, remains the most optimal geometry permitting accurate reconstruction of the ultrafast polarization, current or magnetization dynamics using THz emission spectroscopy.

*Conclusion*

As a summary, we presented a quantitative description of the THz emission experiment, in which the THz generation, propagation and detection are precisely accounted for. We have derived the quantitative transfer function, rigorously and in a fully calibrated manner connecting the observable, THz EO signal, to the quantity of interest, ultrafast polarization, current, or magnetization dynamics in the excited sample serving as a THz emitter. Our modelling results are in very good agreement with our experiments. We have shown that a simple direct line-of-sight THz emission spectrometer geometry is the most optimal one for this kind of measurements. Our simulation indicates that even minor misalignment in a spectrometer with a more complex configuration will lead to the significant signal distortion, which will be impossible to accurately correct for, making the THz signal reconstruction impossible.